\documentclass{article}

\usepackage{arxiv}
\usepackage{lineno}
\usepackage{amsmath}
\usepackage[separate-uncertainty=true]{siunitx}
\usepackage{float}
\usepackage[position=top]{subcaption}
\usepackage{booktabs}
\usepackage{xr}       
\usepackage{hyperref}
\usepackage{graphicx}
\usepackage{tabularx}
\usepackage{natbib}

\title{Resolving multiple layer thicknesses and birefringence in thin films with an optical in-line Roll-to-Roll characterization setup}

\author{S\o{}ren A. R. Kynde\\
	Danish Fundamental Metrology A/S\\
	Kogle All\'{e} 5, H\o{}rsholm, 2970, Denmark\\
		\And
	Astrid T. R\o{}mer\\
	Danish Fundamental Metrology A/S\\
	Kogle All\'{e} 5, H\o{}rsholm, 2970, Denmark\\
		\And
	Abdelouadoud Mammeri\\
	Department of Energy Conversion and Storage, Technical University of Denmark\\
	Fysikvej, Building 310, 2800 Kgs. Lyngby, Denmark\\
		\And
	Jens W. Andreasen\\
	Department of Energy Conversion and Storage, Technical University of Denmark\\
	Fysikvej, Building 310, 2800 Kgs. Lyngby, Denmark\\
		\And
	Moises Espindola\\
	F. Junckers Industrier A/S, V\ae{}rftsvej 4, 4600 K\o{}ge, Denmark\\
		\And
	Matteo Ciambezi\\
	Independent Scientist, Malm\"o{}, Sweden\\
		\And
	S\o{}ren A. Jensen\\
	Kogle All\'{e} 5, H\o{}rsholm, 2970, Denmark\\
	\texttt{saj@dfm.dk}\\
}

\renewcommand{\headeright}{}
\renewcommand{\undertitle}{}

\begin{document}
\maketitle

\begin{abstract} 
Optical techniques provide a fast and flexible path to material inspection. In particular, the investigation of key metrics during material production can help obtain the optimal parameters to achieve desired material functionality as well as to ensure product quality.
Here we present an optical imaging setup designed for fast material inspection in a roll-to-roll deposition environment. The method allows for robust simultaneous determination of thickness in the range from \qtyrange{0.3}{110}{\micro \metre} of multiple layers of transparent or semi-transparent materials in the form of a free film or a film deposited on a thicker transparent substrate. The PEDOT and ZnO materials chosen for this proof of concept are relevant for the fabrication of organic solar cells.  Additionally, the method may be used to characterize birefringence.

\end{abstract}

\section{Introduction}
Printed thin films with thicknesses on the \qty{}{\nano\metre} to \qty{}{\micro\metre} scale find applications in electronics \cite{Chandrasekaran2022}, opto-electronic devices such as photovoltaics \cite{Sun2022,Alharbi2025} and sensors \cite{Maddipatla2020}, but also as membranes in for instance power-to-X electrolyzers and fuel cells \cite{PaixaodaCosta2024}.

Among the available wet-coating processes, slot-die coating (SDC) has proven capable of producing uniform films with low defect densities, even over large areas and at high speeds. 
It is a pre-metered, meniscus-based deposition method in which a liquid precursor is delivered through a narrow slit (the die), with a blade (the meniscus guide) controlling the coating width deposited onto a substrate. Due to its reproducibility, material efficiency, and adaptability to continuous manufacturing processes, SDC has gained widespread adoption in both research and industrial settings for the fabrication of functional coatings \cite{li_recent_2020, park_roll_roll_2020}.

Measurement of thin film thickness using spectral data in the visible or near-visible range has been demonstrated before, see e.g.\cite{Kubinyi1996, Nevas2003}. Where previous results are based on single-point measurements of non-moving samples, the present work implements a similar analysis to data from a line scan hyperspectral camera. Digital image processing of the line scan data in combination with the translation of the sample, as it naturally occurs in a roll-to-roll manufacturing process, enables us to generate a thickness map of the supporting substrate and the layer being deposited onto it. In the direction of translation, the resolution of the map depends on the processing speed of the analysis relative to the translation speed of the film. In the transverse direction, it depends on the number of pixels in the camera relative to the width of the observed area. 

Here we demonstrate the method on a small (\qty{20}{\milli\metre} wide) area with a resolution of \qty{80}{\micro\metre}. By adapting the moving speed and choosing suitable optics, the method can easily be extended to a larger field of view with a correspondingly lower spatial resolution.

\section{Methods}
\label{sec:methods}
The hyperspectral camera method for real-time in-line measurements in a roll-to-roll fabrication environment is described in Ref.\cite{SkovlundMadsen2021}, where it was used to characterize periodic nanostructures fabricated by nano-embossing. In short, collimated white light from a fiber-coupled light source is transmitted through the sample film. The transmitted light is measured by a hyperspectral linescan camera. For the measurements presented in sections \ref{sec:offline} and \ref{sec:inline} we used a hyperspectral camera setup consisting of a fast monochrome camera (Andor Zyla 5.5) and a spectrograph (Andor Kymera-193i) in combination with a Xe-lamp light source, which has high intensity in the visible range. This setup records the transmission spectrum in the wavelength range \qtyrange{450}{720}{\nano\metre} with sub-nm resolution. For the measurements presented in section~\ref{sec:oxidethinfilms} we used a Resonon Pika XC2 hyperspectral camera system in combination with a halogen light source, which provides visible and near infrared light. This system records the spectrum at wavelengths from \qtyrange{400}{1000}{\nano\metre} with a spectral resolution of about \qty{1.7}{\nano \metre}.

The recorded image is one-dimensional in its spatial extent, but each spatial point or \emph{track} has the full spectral information generated by the spectrograph. Orienting the one-dimensional spatial viewing direction perpendicular to the movement direction in a roll-to-roll fabrication environment, we can record consecutive \emph{frames} and bring them together into a two-dimensional spatial image where each spatial point has the full spectral information. 

In Ref. \cite{Madsen2022}, we demonstrated how this measurement reveals the thicknesses of transparent thin films.
Transmission of broadband visible light through semitransparent media is resolved in its wavelength components, displaying an oscillating pattern caused by optical interference stemming from internal reflections in the material. The characteristic frequency of the oscillation is determined by the optical path length ($L = n \cdot d$) through the material, where $n$ is the refractive index of the material and $d$ is the thickness. In this way, the thickness of 
a layer of material is related to the characteristic frequency. 

\begin{figure}
\centering
\subfloat[]{\includegraphics[width=0.48\textwidth]{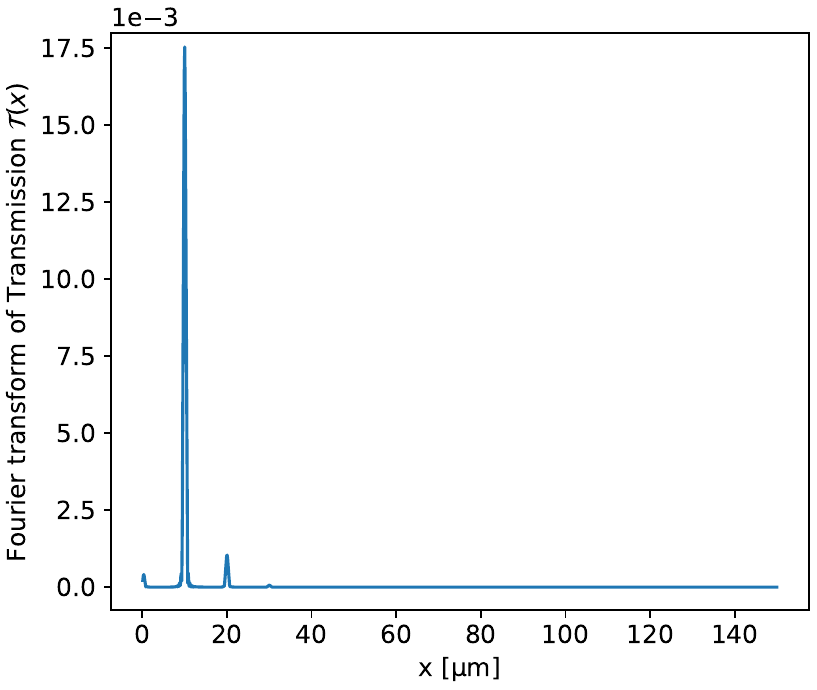}}
\quad
\subfloat[]{\includegraphics[width=0.48\textwidth]{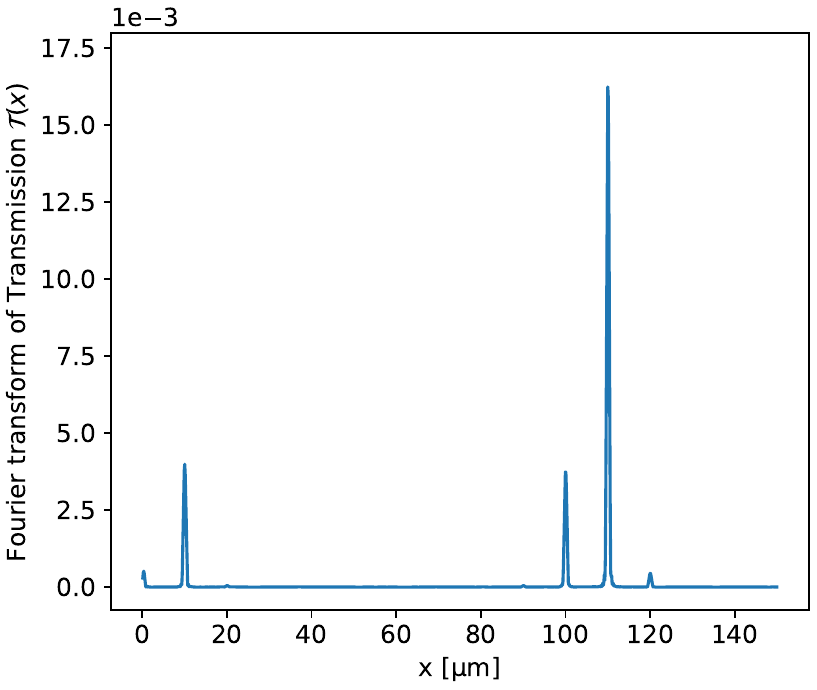}}
\caption{(a) Simulated transmission spectrum of a \qty{10}{\micro \metre} thick PET film calculated using Eq.\eqref{eq:T}.
(b) Simulated transmission spectrum for a sample consisting of one \qty{10}{\micro\metre} PET layer, a \qty{10}{\nano\metre} air layer to simulate an imperfect interphase, and an additional \qty{100}{\micro\metre} thick PET layer, calculated using the transfer-matrix method \cite{byrnes2020multilayeropticalcalculations}.}
\label{fig:Film_sim}
\end{figure}

We apply a Fourier transformation approach to extract the thickness, utilizing that 
the expected form of the transmittance through a single layer is given by \cite{hecht2017optics}:
\begin{equation}
T(\lambda)=t_{12}^2t_{23}^2\left[ 1+r_{21}^2r_{23}^2-2r_{21}r_{23}\cos\left(2\pi \tfrac{2n(\lambda)}{\lambda} d\right) \right]^{-1}
\label{eq:T}
\end{equation}
where $t_{ij}$ and $r_{ij}$ are the Fresnel transmission and reflection coefficients between the medium $i$ and $j$, where medium 1 is air, medium 2 is the film, and medium 3 is air.
Knowing the wavelength dependence of the refractive index, $n(\lambda)$, we transform the data by interpolating to an equidistant vector of a new variable $\beta=\frac{2 n(\lambda)}{\lambda}$. 
The intensity $T(\beta)$  oscillates with a frequency of $d$ and its Fourier transform $\mathcal{T}(x)$ will exhibit a characteristic peak at $x=d$, as illustrated in the simulated spectrum shown in Fig.~\ref{fig:Film_sim}~(a). In addition, higher order peaks corresponding to light reflecting multiple times in the layer before exiting on the side facing the camera are present. However, these higher order peaks are effectively reduced at each round-trip owing to the small values of $r_{21}$ and $r_{23}$, which are on the order of 0.24 for a PET film in air. In our experiments, the higher order peaks are therefore not visible.

In the method described above, we underline the importance of knowing the wavelength-dependent refractive index of the material under investigation in order to obtain the correct thickness.
Interestingly, the method allows for a simultaneous measurement of the thicknesses of several material layers inside the film, if these display optically resolvable interfaces. 
The multiple layers will show up as individual peaks in the Fourier spectrum, as well as one or more sum peaks corresponding to the optical path length through multiple layers, as shown in the simulated spectrum in Fig.~\ref{fig:Film_sim}~(b).

The wavelength sampling must be sufficiently fine to resolve the thickness of the film. If we are to detect a sample of thickness $d$, the sampling steps must be at least $\beta=\frac{1}{2d}$, according to the sampling theorem. The inverse relationship between the parameter $\beta$ and the wavelength $\lambda$ results in a closer sampling of $\beta$ in the long-wavelength part of the spectrum, assuming a constant wavelength sampling resolution. 
Using the Andor Zyla camera setup with the high spectral resolution of \qty{0.5}{\nano\metre} in the wavelength range \qtyrange{450}{720}{\nano\metre}
region, we can resolve film thicknesses up to values 
over \qty{100}{\micro\metre}, depending on the material. With the lower spectral resolution of roughly \qty{1.7}{\nano\metre} achievable with the other camera system, we can obtain a similar sensitivity for thick films when the recorded spectrum includes wavelengths up to \qty{1000}{\nano\metre}.

With regard to the sensitivity towards very thin films, a wide spectrum of wavelengths becomes essential. 
In this case, the measurement sensitivity is limited by the ability to detect a full oscillation of the interference pattern. As we show in section~\ref{sec:oxidethinfilms}, we can detect oxide film thicknesses down to \qty{340}{\nano\metre} when invoking the camera solution that detects a wavelength range of \qtyrange{400}{1000}{\nano\metre}.

Thin-film deposition was carried out using a roll-to-roll coating system (FOM Technologies SigmaR2R). The precursor solution was delivered by a syringe pump (Chemyx) to a stainless-steel slot-die head (FOM Technologies) with a coating width of \qty{13}{\milli \metre}, as determined by the meniscus guide. 

\section{Measurements and analysis}

\subsection{Offline measurements}
\label{sec:offline}

\begin{figure}
\centering
\subfloat[]{\includegraphics[height=0.3\textwidth]{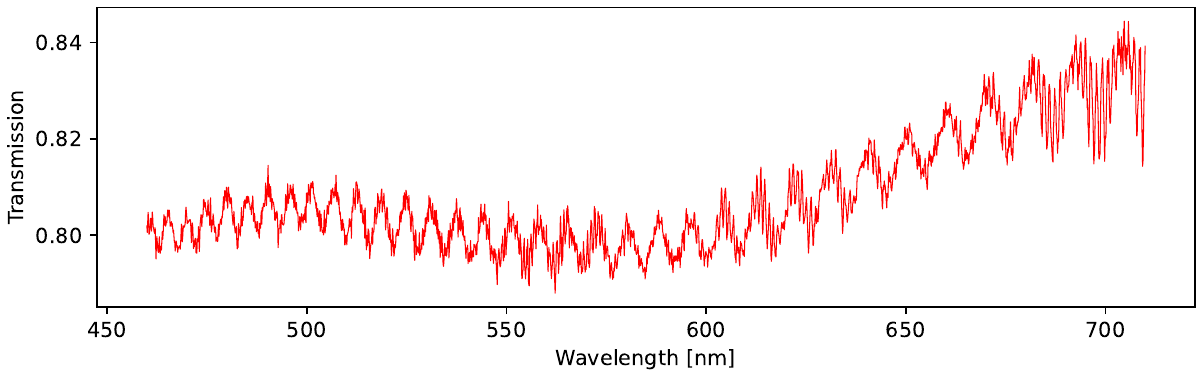}}
\\
\subfloat[]{\includegraphics[height=0.3\textwidth]{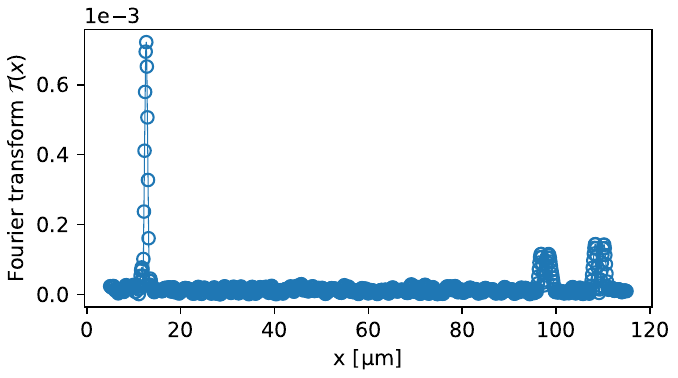}}
\subfloat[]{\includegraphics[height=0.3\textwidth]{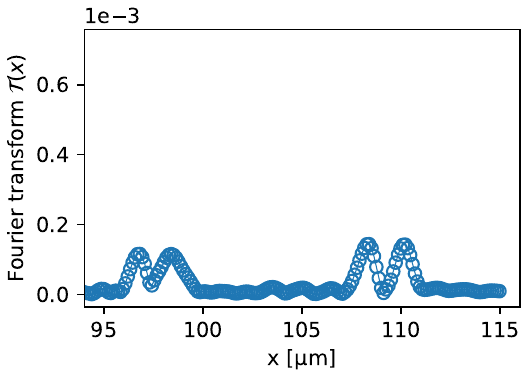}}
\caption{Thickness extraction on Melinex PET double layer film using unpolarized light. (a) Raw transmission spectrum as a function of wavelength ($\lambda$). (b) The transmission spectrum was transformed to express transmission as a function of $2n(\lambda)/\lambda$ and then Fourier transformed. The Fourier transform reveals three peaks at characteristic path lengths $x=d_1,d_2,d_3$ corresponding to a thin adhesive layer ($d_1=\qty{12.6}{\micro\metre}$), a thick PET substrate layer ($d_2=\qty{96.8}{\micro\metre}$) and the sum of both layers ($d_3=\qty{109}{\micro\metre}$). (c) Zoom-in on the peak for the thick substrate layer and the sum, revealing that they are both double peaks, owing to the birefringence in the PET substrate.}
\label{fig:Melinex_unpol}
\end{figure}

\begin{figure}
\centering
\subfloat[]{\includegraphics[height=0.3\textwidth]
{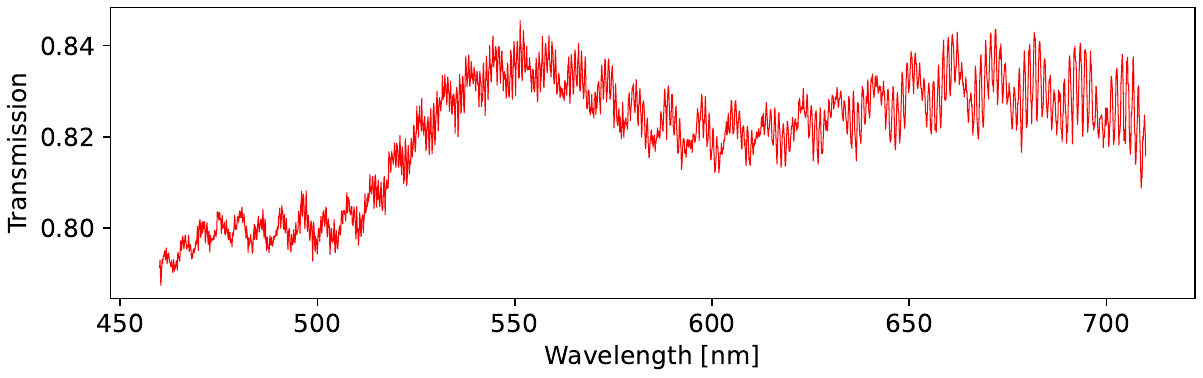}}
\\
\subfloat[]{\includegraphics[height=0.3\textwidth]
{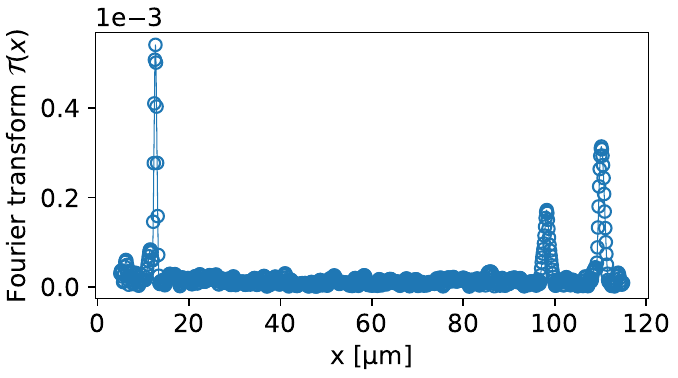}}
\subfloat[]{\includegraphics[height=0.3\textwidth]
{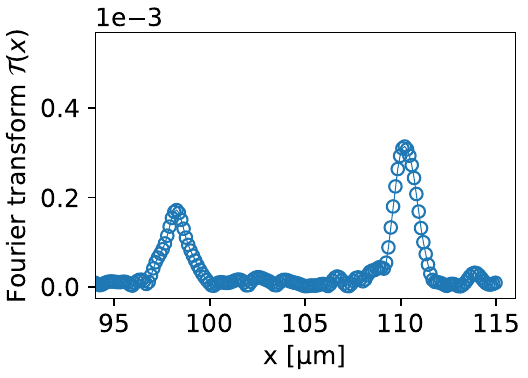}}
\caption{Thickness extraction on Melinex PET double layer film using polarized light. (a) Raw transmission spectrum. (b) Fourier transform of data in (a) expressed as a function of $2n(\lambda)/\lambda$, revealing three characteristic peaks corresponding to a thin layer, a thick layer and the sum of both layers. (c) Zoom-in on the peak for the thick substrate layer and the sum, showing only single peaks.
}
\label{fig:Melinex_pol}
\end{figure}

First, we analyze a simple test sample consisting of a PET film with a thin layer of amorphous PET on top. A transmission spectrum from a \qty{0.4}{\milli \metre} x \qty{0.4}{\milli \metre} area of this sample is shown in Fig.~\ref{fig:Melinex_unpol}(a). It is possible to discern different characteristic frequencies in the transmission spectrum. This becomes even more clear in the Fourier transform in Fig.~\ref{fig:Melinex_unpol}(b), where the corresponding frequency peaks show up. Three peaks are seen in Fig.~\ref{fig:Melinex_unpol}(b), one at \qty{12.6}{\micro\metre} corresponding to the thin amorphous layer, one at \qty{96.8}{\micro\metre} corresponding to the substrate film, and one at \qty{109}{\micro\metre} corresponding to the combined thickness of the two layers. The total thickness of the film could readily be confirmed with a handheld calibrated micrometer-screw, which yielded a thickness of \qty{109}{\micro\metre} with an estimated expanded measurement uncertainty (\qty{95}{\percent} confidence) of \qty{2}{\micro\metre}. Fig.~\ref{fig:Melinex_unpol}(c) shows a zoom-in on the two peaks at large thickness. We note that each peak is in fact a double peak, which we attribute to birefringence effects of the PET substrate layer: the refractive index takes different values along different axes of the material, and this in turn gives rise to different optical path lengths, amounting to the appearance of different thicknesses. This is consistent with the observation that the peak corresponding to the thin amorphous film does not show the double peak behavior, as this film layer is not birefringent. Fig.~\ref{fig:Melinex_pol} shows a similar measurement on the same sample, but with a polarizer in the incoming light beam oriented in the direction corresponding to the highest index of refraction in the material. This was found to coincide with the roll direction of the foil. Introducing the polarizer eliminates the left twin of each of the double peaks in the transmission spectrum.

\begin{figure}
\centering
\subfloat[]{\includegraphics[width=0.48\textwidth]{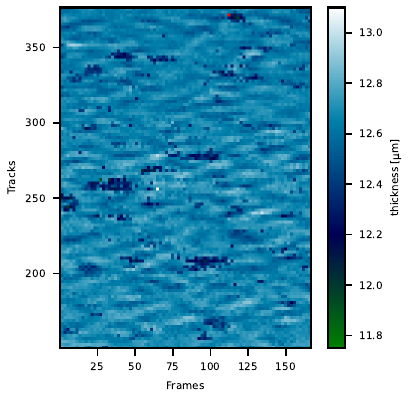}}
\quad
\subfloat[]{\includegraphics[width=0.48\textwidth]{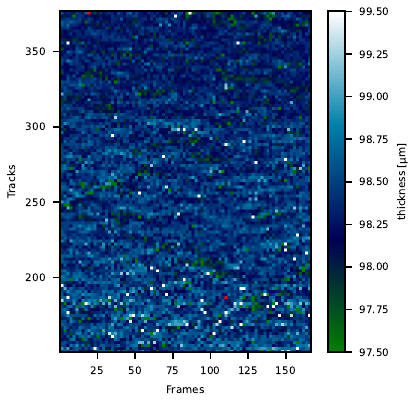}}
\caption{Thickness maps on the Melinex PET double layer film measured with polarized light produced by performing the analysis shown in Fig.~\ref{fig:Melinex_pol} in small areas (2 frames $\times$ 2 tracks) to find the thickness of the thin layer (a) and the thick layer (b). Red dots represent areas where a fit to a peak in the Fourier spectrum could not be performed. Note, for some tracks the field of view was obscured by the polarizer, these are not shown in the maps.}
\label{fig:Melinex_maps}
\end{figure}

By using the positions of the remaining peaks with a polarizer in the incoming light beam as the best estimate of the layer thicknesses, we can generate the two-dimensional thickness maps shown in Fig.~\ref{fig:Melinex_maps}. These were obtained by moving the sample relative to the hyperspectral camera while measuring, and performing the thickness analysis in each point. In Fig.~\ref{fig:Melinex_maps}, both the thickness of the thin layer (a) and the thick substrate (b) were derived simultaneously from a single measurement at each of the tracks. The axis denoted 'tracks' corresponds to one row of pixels in the camera. The axis denoted 'frames' is the result of continuous measurements of all the 'tracks' as the sample moves by the camera. In this particular example 550 tracks were focused on a \qty{20}{\milli \metre} illuminated strip of the sample and 175 frames were captured during a \qty{7}{\milli \metre} long movement of the sample. Thus each 'pixel' of the raw data corresponds to approx. $\qty{0.04}{\milli \metre} \times \qty{0.04}{\milli \metre}$. In the presented map, the tracks and frames were analyzed in a $2 \times 2$ binning to improve data quality, resulting in an array of $\qty{0.08}{\milli \metre} \times  \qty{0.08}{\milli \metre}$ squares. The spatial resolution can easily be adapted to a variety of situations by altering the cameras field of view, the speed of movement, or the camera frame rate, and by altering the binning.

The average thickness of the thin layer from the data shown in Fig.~\ref{fig:Melinex_maps} (a) is \qty{12.6}{\micro\metre} with a standard deviation of \qty{0.1}{\micro\metre}, and the average thickness of the thick layer in Fig.~\ref{fig:Melinex_maps} (b) is \qty{98.4}{\micro\metre} with a standard deviation of \qty{0.4}{\micro\metre}. The average total thickness is thus \qty{111.0}{\micro\metre}, which is consistent with the result of \qty{109(2)}{\micro\metre} from the micrometer-screw.

\subsection{In-line measurements}
\label{sec:inline}

\begin{figure}
\centering
\subfloat[]{\includegraphics[width=0.48\textwidth]{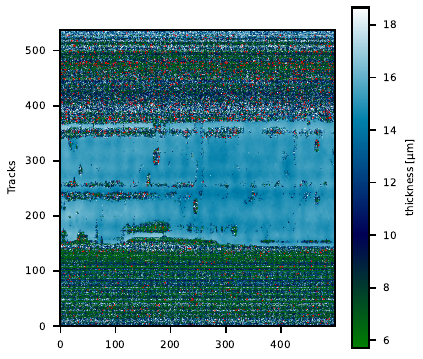}}
\mbox{}\hfill
\subfloat[]{\includegraphics[width=0.48\textwidth]{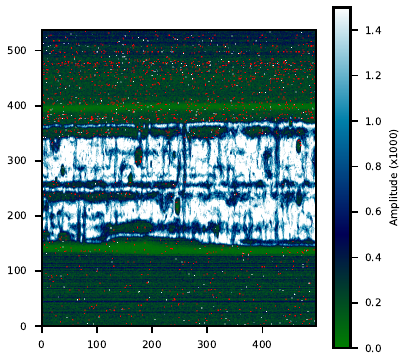
}}
\\
\subfloat[]{\includegraphics[width=0.48\textwidth]{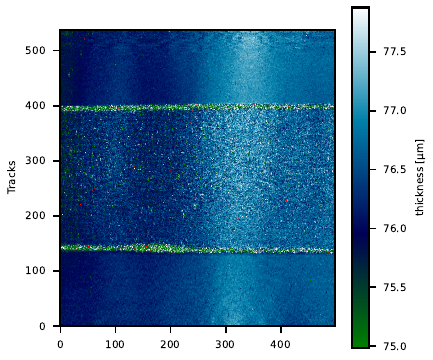}}
\mbox{}\hfill
\subfloat[]{\includegraphics[width=0.48\textwidth]{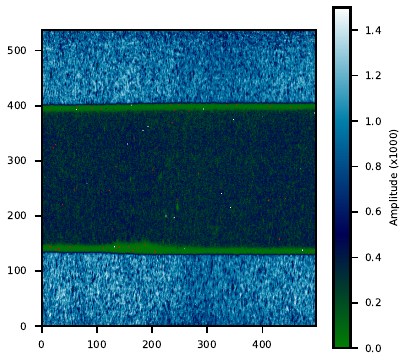}}
\\
\subfloat[]{\includegraphics[width=0.48\textwidth]{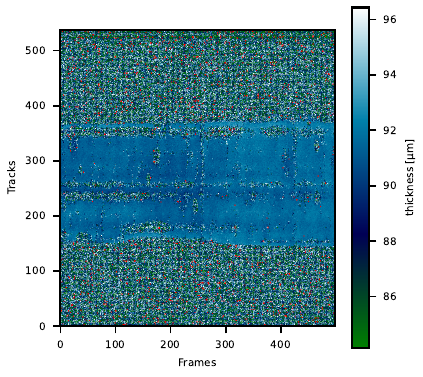}}
\mbox{}\hfill
\subfloat[]{\includegraphics[width=0.48\textwidth]{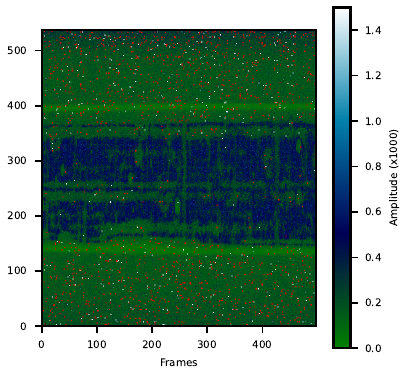}}
\caption{Thickness mapping on PEDOT:PSS wet film just after deposition on a PET foil. Left column: Position of thickness peak for the thin peak corresponding to the deposited layer (a), the substrate (c) and the sum of the two (e). Right column: Amplitude of the thickness peaks in (a), (c) and (e) respectively.}
\label{fig:PEDOT_FET}
\end{figure}

The hyperspectral camera system was tested in-situ on the roll-to-roll slot die coater from FOM Technologies described in section~\ref{sec:methods}, here measurements could be performed during layer deposition with the camera system mounted just after the slot-die head. Fig.~\ref{fig:PEDOT_FET} shows the result of a thickness analysis on a \qty{13}{\milli\metre} wide layer of PEDOT:PSS (Heraeus CLEVIOS FET) being printed on a PET substrate and measured on the wet film just after deposition.
 The film was moving along the horizontal direction in the maps, denoted by Frames, and the layer was only printed on the region in the middle of the measured area. The measurement, which is shown in Fig.~\ref{fig:PEDOT_FET}, was performed with a frame rate of \qty{96.5}{fps} on a film being deposited with a roll speed of \qty{0.6}{\metre \per \min}, resulting in a spacing between recorded frames of \qty{0.1}{\milli \metre}.
Similar to the double layer film studied in the previous section, there were three visible peaks in the Fourier transform of the transmission spectrum, corresponding to a thin deposited layer, a thicker substrate film, and the combined thickness. Fig.~\ref{fig:PEDOT_FET}(a), (c), and (e) show maps of the derived thicknesses for the thin deposited layer, thicker substrate layer, and sum, respectively. Here we have assumed the refractive index of PET for all materials. It is seen in Fig.~\ref{fig:PEDOT_FET}(a) that the thin layer is well-defined in the middle area where the film is deposited, apart from a few patches where the thickness is poorly defined and the peak amplitude is low, Fig.~\ref{fig:PEDOT_FET}(a) and (b) respectively, presumably because of imperfect deposition causing an uneven film surface.
 On the other hand, the substrate thickness shown in Fig.~\ref{fig:PEDOT_FET}(c) is present everywhere, but most clear at the edges where no film is printed. The sum peak in Fig.~\ref{fig:PEDOT_FET}(e) is clear in the same areas as the printed film, which is to be expected as this peak would only show up when there is a printed layer. Figure~\ref{fig:PEDOT_FET} (b), (d), and (f) show the respective amplitudes of the three thickness peaks in the Fourier transmission, confirming that the thin layer peak and the sum peak are seen in the middle area where a layer is printed, and the substrate peak is strongest at the edges. Omitting the regions where the peak amplitude is low, we find that the deposited PEDOT layer has an average thickness of \qty{15.1}{\micro \metre} with a standard deviation of \qty{0.4}{\micro \metre}, comparable to the estimated thickness from the deposition conditions which is about \qty{13}{\micro \metre}.

The computation time for each pixel depends on the number of sampling points in the wavelength spectrum. With our implementation of a Fourier transform (Fast Fourier Transform - FFT) plus peak fitting to the spectra we observe a more or less linear dependency on the number of sampling points, indicating that the fitting is the dominant contribution to the computation. On a 4-kernel desktop computer with 3 GHz processor the computations were done at \qty{3}{\milli \second} per spectrum of 600 sampling points. Thus, each frame of 500 tracks could be analyzed 1.5 seconds. To be able to analyze at the same rate as the data are recorded would require a higher computing capability, a narrowing of the analyzed tracks and/or a reduction of the spatial resolution by e.g. a factor of ten in each direction. Note that the computations are well suited for parallelization since each pixel is analyzed separately.  

\subsection{Oxide Thin films}
\label{sec:oxidethinfilms}
While resolving relatively thick transparent layers in the \qty{100}{\micro\metre} range requires a good spectral resolution, the ability to measure very thin film layers is much less sensitive to the spectral resolution. On the other hand, it demands a broad wavelength spectrum. To explore thin film layers, we invoke the Resonon Pika XC2 hyperspectral camera system with the wavelength range of \qtyrange{400}{1000}{\nano\metre}. 
The Fourier transformation method relies on having at least one full period completed within the detected spectral range. In general, we expect that a spectrum which includes more than one period will result in a more accurate thickness measurement. 
Below, we explore the results of using the methods to determine oxide film thicknesses below \qty{1}{\micro \metre}.

\begin{figure}
\centering
\subfloat[]{\includegraphics[width=0.28\textwidth]{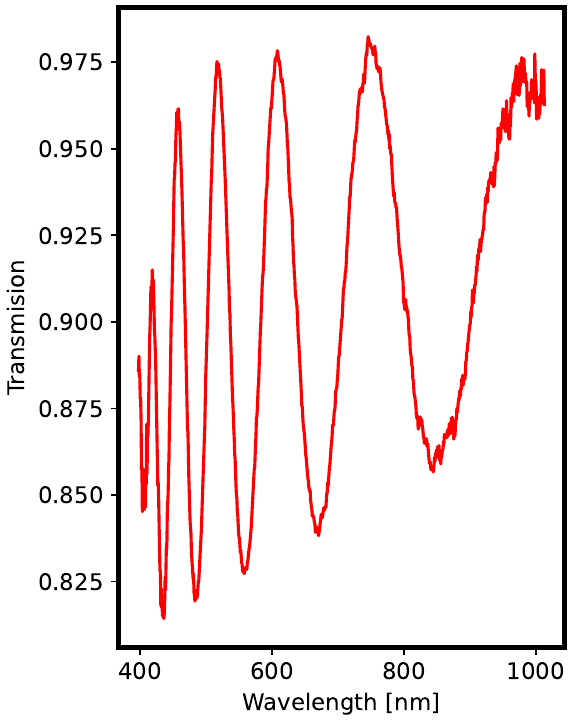}}
\mbox{}\hfill
\subfloat[]
{\includegraphics[width=0.28\textwidth]{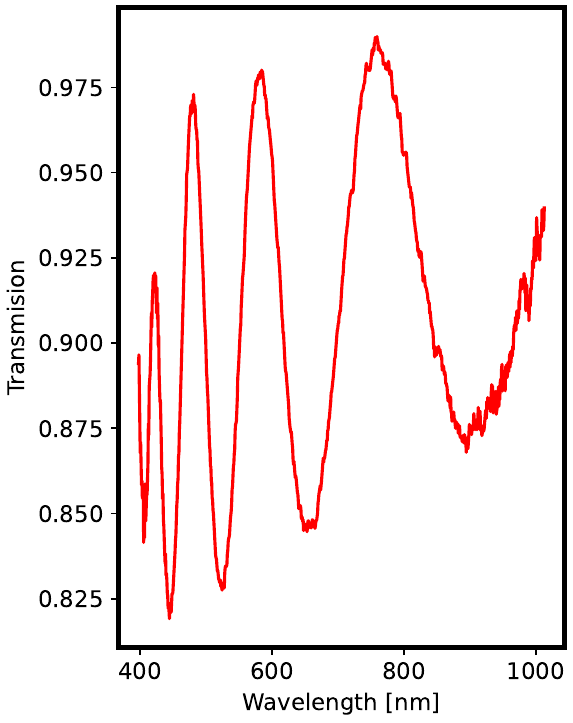}}
\mbox{}\hfill
\subfloat[]
{\includegraphics[width=0.28\textwidth]{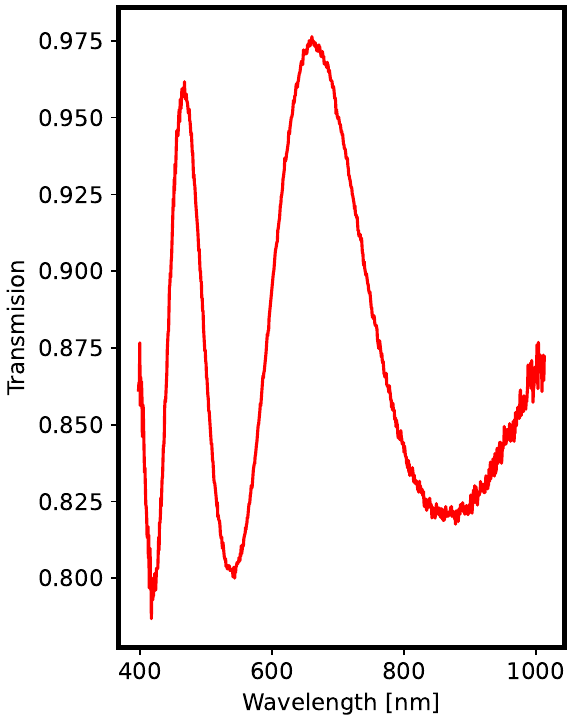}}
\\
\subfloat[]
{\includegraphics[width=0.28\textwidth]{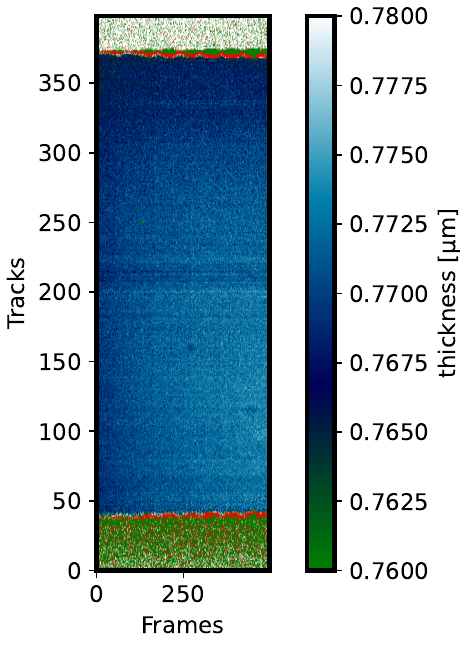}}
\mbox{}\hfill
\subfloat[]
{\includegraphics[width=0.28\textwidth]{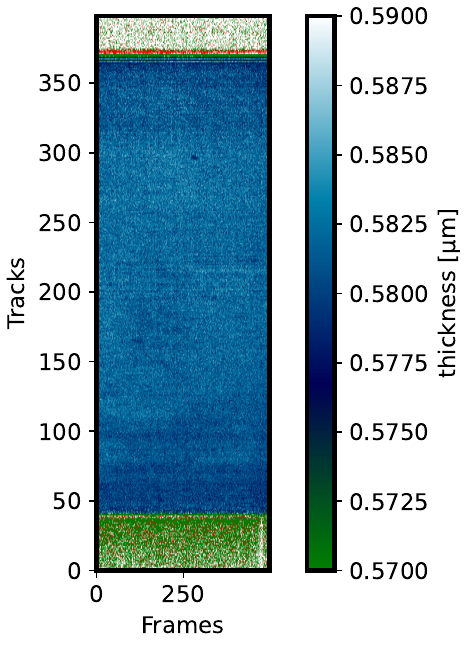}}
\mbox{}\hfill
\subfloat[]
{\includegraphics[width=0.28\textwidth]{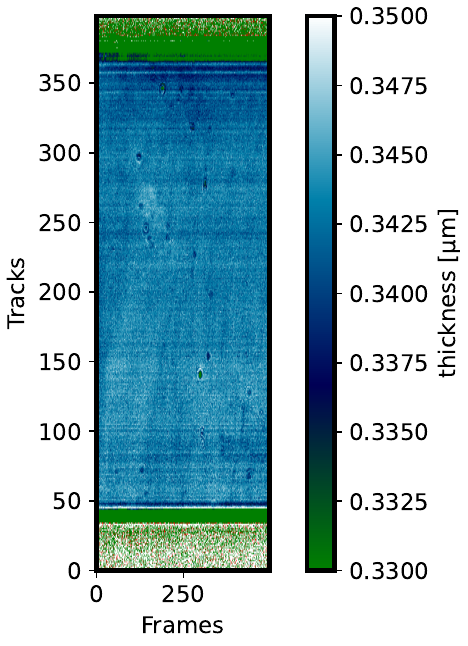}}
\\
\subfloat[]{\includegraphics[width=0.28\textwidth]{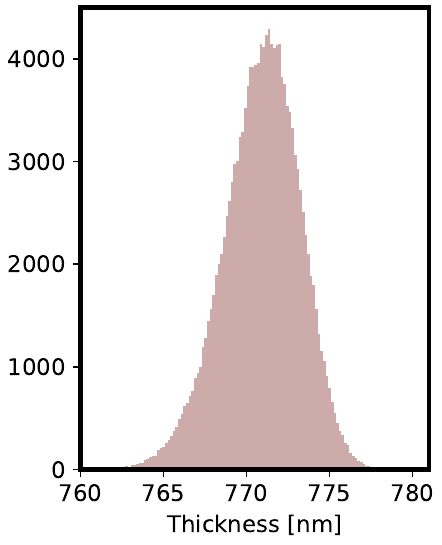}}
\mbox{}\hfill
\subfloat[]{\includegraphics[width=0.28\textwidth]{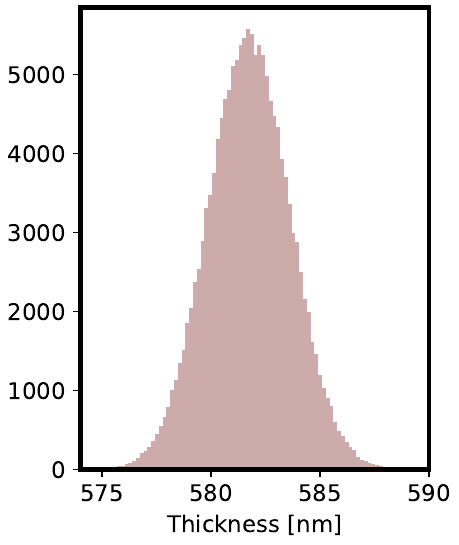}}
\mbox{}\hfill
\subfloat[]
{\includegraphics[width=0.28\textwidth]{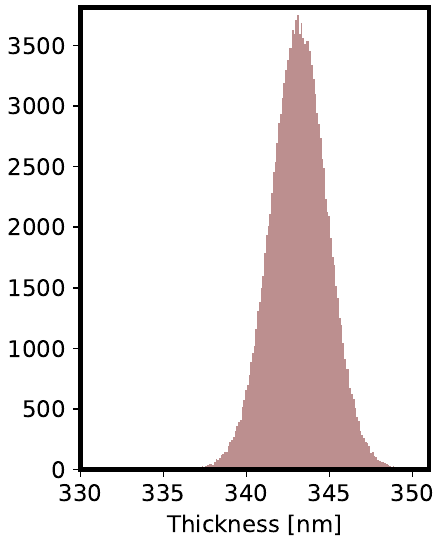}}
\caption{(a,b,c) Raw transmission spectrum of thin zinc-oxide films deposited on glass and measured with the Resonon camera for a spectral range of \qtyrange{400}{1000}{\nano\metre}. (d,e,f) Thickness mapping of the oxide layers for each of the above spectra. Note that the colorbar has been adjusted in each case to highlight the small variations in deposited layer thickness. (g,h,i) Histograms of the thickness of the deposited zinc-oxide layers in each of the three cases. 
}
\label{fig:ZnO_Resonon}
\end{figure}

Figure~\ref{fig:ZnO_Resonon}(a-c) show raw transmission measurements for the thickness determination of thin zinc-oxide (ZnO) layers sputtered on glass (experimental details in the supplementary table 2). 
We investigate films of three different thicknesses ($d_1$, $d_2$ and $d_3$). For all measurements, we invoke the method of Fourier transformation to estimate the thickness of the ZnO layers using a binning of 4 pixels along the Track direction and a refractive index given from ellipsometry measurements of each film separately (data and model in supplementary figures 1, 2, 3, 4 and table 1). 
Figure~\ref{fig:ZnO_Resonon}(d-f) show the measured film thicknesses obtained by the Fourier analysis, showing that the deposited layers are relatively uniform. 
To further examine the uniformity of the thin films, we plot the histograms of the film thicknesses in the range Track $=400-1200$ for the three films in Fig.~\ref{fig:ZnO_Resonon}(g-i). The data are, to a good approximation, described by a Gaussian distribution, as expected since the films were deposited by sputtering under rotation. 
The mean values and Gaussian widths of the thickness distributions are listed in Table~\ref{tab:ZnO}. Also shown are the thickness values measured by ellipsometry, using the same refractive index data for the three films as used in the analysis of the camera data. These values agree reasonably well with the values measured by the camera system.

\begin{table}
\centering
\caption{Average thickness values for the three films presented in Fig.~\ref{fig:ZnO_Resonon}. Also shown are the Gaussian widths ($\sigma$) of the thickness histograms in Fig.~\ref{fig:ZnO_Resonon} and thickness values measured with ellipsometry on the same films.}
\begin{tabularx}{\textwidth}{ c c c c }
\toprule
Film&Extracted thickness [nm] & $\sigma$ value [nm] & Thickness from ellipsometry [nm]\\
\midrule
$d_1$&343 & 1.8 & 338\\
$d_2$&582 & 1.9 & 586\\
$d_3$&771 & 3.9 & 774\\

\bottomrule
\end{tabularx}
\label{tab:ZnO}
\end{table}

\section{Conclusion}

We present an in-line characterization tool for real time areal measurements of  film thickness during a roll-to-roll film deposition process. The tool is based on a hyperspectral line scan camera, providing full imaging of layer thickness during deposition, but it can very well be used at any point of a fabrication process. We show that multiple layer thicknesses can be extracted from a single measurement, making it possible to perform simultaneous characterization on multi-layer samples. Layer thicknesses ranging from \qty{0.34}{\micro\metre} to \qty{110}{\micro\metre} were successfully extracted.

In this method, materials exhibiting birefringence will appear to have multiple thicknesses. This can be remedied by using light polarized in one of the axes of birefringence.
We envision that the effect can be exploited to analysis birefringent materials either by using unpolarized light to map the splitting of the peaks as a measure of the extent of birefringence or by using several scans of varied polarization to map the local orientation of the birefringence axes of a material.

\section*{Acknowledgments}
The authors thank FOM Technologies A/S for supporting the experiments and providing access to slot-die coating equipment.

\section*{Funding}
This work was partly funded by Innovation Fund Denmark (IFD), Eureka Eurostars project E2897 QualSurf. The project is supported by funds from the Danish Agency for Higher Education and Science.

\section*{Disclosures}
The authors declare that there are no conflicts of interest related to this article.

\section*{Data availability}
Data presented in this paper may be obtained from the authors upon request.


\bibliographystyle{unsrt}
\bibliography{References}

\end{document}


\title{Resolving multiple layer thicknesses and birefringence in thin films with an optical in-line Roll-to-Roll characterization setup: Supplementary Material}
\maketitle

\section*{Ellipsometry}
\begin{figure}[!ht]
    \centering
    \includegraphics[width=1\linewidth]{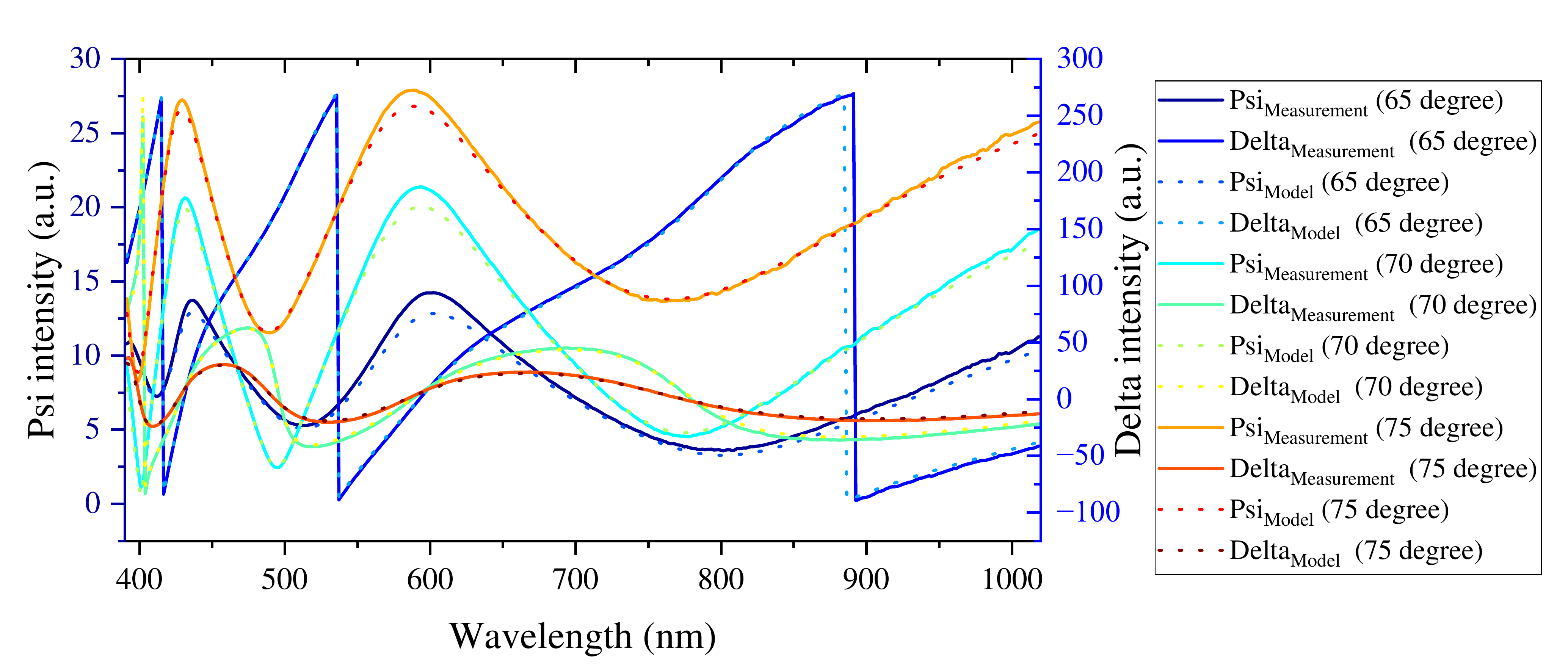}
    \caption{Ellipsometry measurements and fitted model at 65, 70 and 75 degrees angles for the $d_1$ sample with mean square error of 11.9.}
    \label{ellip_340}
\end{figure}
\begin{figure}[!ht]
    \centering
    \includegraphics[width=1\linewidth]{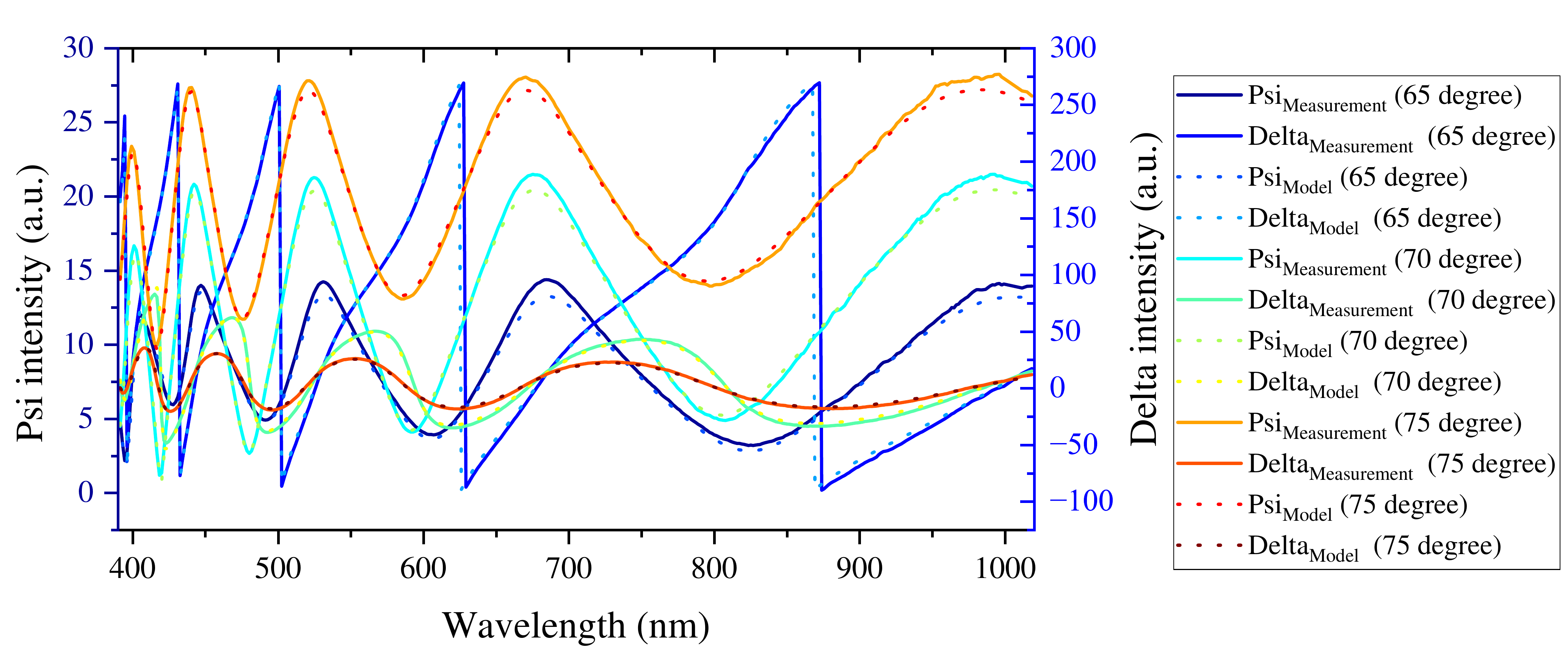}
    \caption{Ellipsometry measurements and fitted model at 65, 70 and 75 degrees angles for the $d_2$ sample with mean square error of 12.1.}
    \label{ellip_500}
\end{figure}
\begin{figure}[!ht]
    \centering
    \includegraphics[width=1\linewidth]{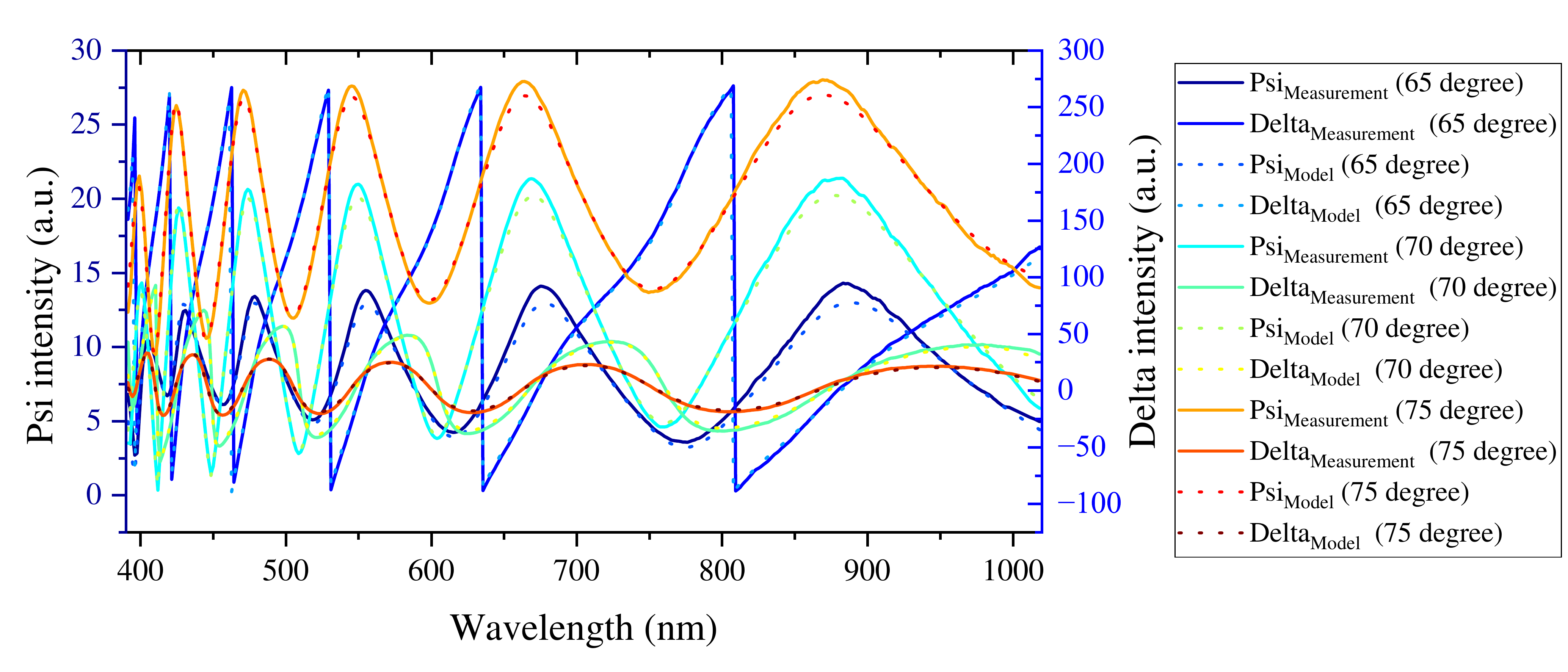}
    \caption{Ellipsometry measurements and fitted model at 65, 70 and 75 degrees angles for the $d_3$ sample with mean square error of 12.8.}
    \label{ellip_170}
\end{figure}
Figures \ref{ellip_340}, \ref{ellip_500} and \ref{ellip_170} represent the ellipsometry measured and the fitted data in the interval  \qtyrange{391}{1018}{\nano\metre}. To model the refractive index of ZnO, we fitted the absorption part using the General Oscillator (Gen-Osc) dispersion model, which has a Lorentzian and a Gaussian peak. Then apply the Kramers–Kronig relationship to obtain the real part with certain constraints on $\epsilon_{real}$. The parameters of the model and the values of the optical constants are reported next:
\begin{figure}[!ht]
    \centering
    \includegraphics[width=1\linewidth]{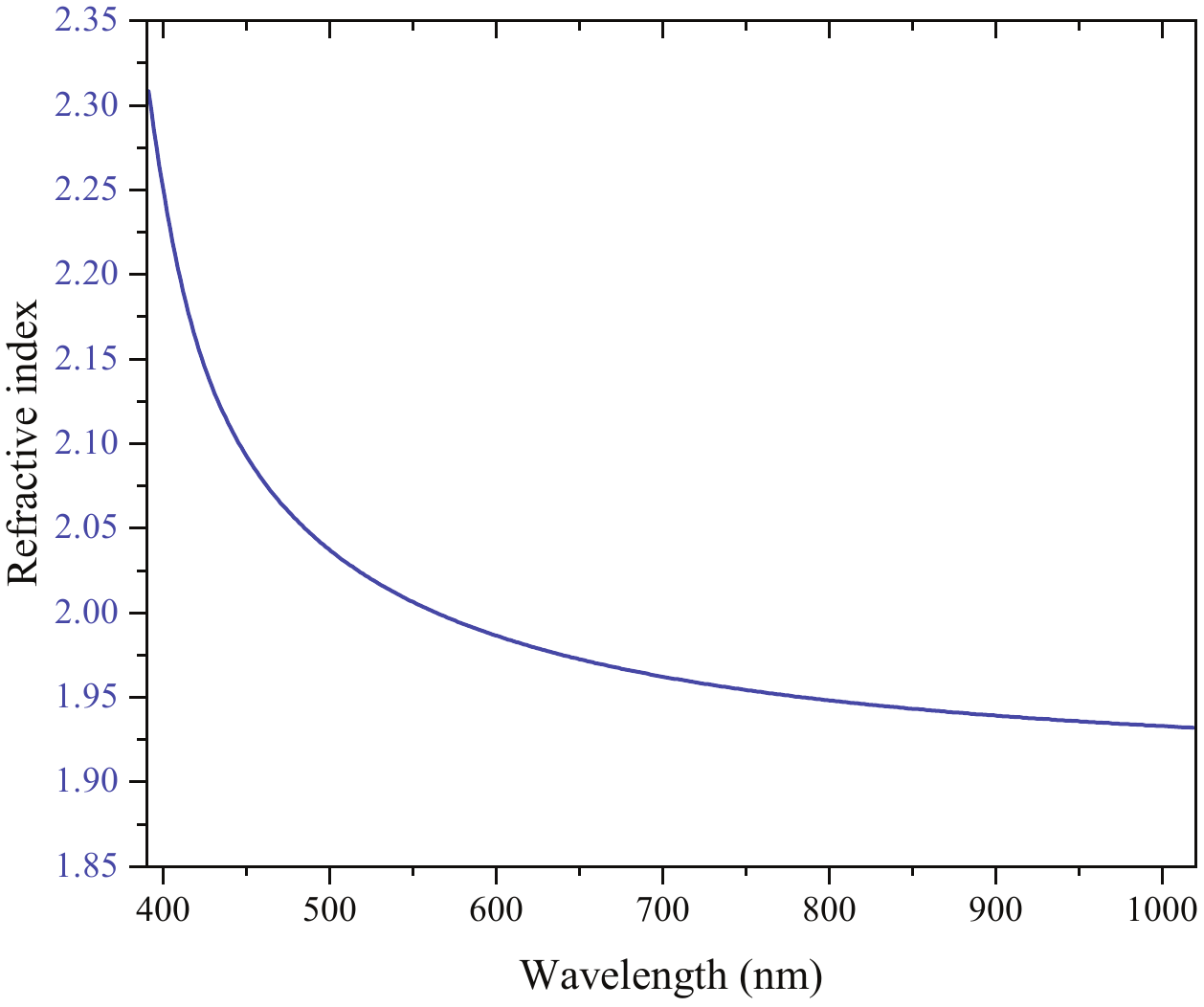}
    \caption{The optical constant of the ZnO sputtered films.}
    \label{n_k}
\end{figure}
\begin{table}[!ht]
\centering
\begin{tabular}{|l|c|}
\hline
\textbf{Parameter} & \textbf{Value} \\
\hline
$E_{\infty}$ & 1.900 \\
UV Pole Amp & 30.5000 \\
UV Pole En (\qty{}{\electronvolt}) & 7.800 \\
\hline
\multicolumn{2}{|c|}{\textbf{Lorentz Oscillator}} \\
\hline
Amplitude $A_1$ & 9.40392066E2 \\
Broadening $Br_1$ (\qty{}{\electronvolt}) & 0.006283 \\
Energy $E_{n1}$ (\qty{}{\electronvolt}) & 6.020 \\
\hline
\multicolumn{2}{|c|}{\textbf{Gaussian Oscillator}} \\
\hline
Amplitude $A_2$ & 3.140451 \\
Broadening $Br_2$ (\qty{}{\electronvolt}) & 0.4410 \\
Energy $E_{n2}$ (\qty{}{\electronvolt}) & 3.598 \\
\hline
\multicolumn{2}{|c|}{\textbf{Surface Roughness (\qty{}{\nano\meter})}} \\
\hline
$d_1$ & 5.69 \\
$d_2$ & 5.93 \\
$d_3$ & 7.74 \\
\hline
\end{tabular}
\caption{Dispersion model parameters of $d_1$, $d_2$, and $d_3$.}
\label{ZnO_ellip_parameters}
\end{table}
We have fitted over the thickness (reported in main text) and the roughness, reported in table \ref{ZnO_ellip_parameters}. 
\section*{ZnO films deposition}
 The ZnO films were deposited using a Lesker CMS sputtering system with a ZnO target (Purity: 99.999\%, Dimension: 2.00" Diameter $\times$ 0.250" Thick and Density: \qty{5.25}{\gram\per\cubic\centi\meter}). Polished 4-inch silica wafers with a thickness of 1000 $\pm$ \qty{25}{\micro\metre} served as substrates (Active Business Company GmbH). Argon  ($99\%$ purity) was used as the sputtering gas, and the substrates were maintained at room temperature under continuous rotation. The rest of the coating parameters for each film are listed in Table \ref{sputtering_parameters}:

\begin{table}[!ht]
\centering
\begin{tabular}{|c|c|c|c|}
\hline
Film & Deposition time (\qty{}{\second}) & Operational pressure (\qty{}{\milli\bar})] & Power (\qty{}{\watt}) \\
\hline
$d_1$ & 17500 & 2.62 & 60 \\
\hline
$d_2$ & 30000 & 3.01 & 60 \\
\hline
$d_3$ & 30000 & 3.02 & 100  \\
\hline
\end{tabular}
\caption{Sputtering parameter for the ZnO films.}
\label{sputtering_parameters}
\end{table}